# Coupled Charge and Radiation Transport Processes in Thermophotovoltaic and Thermoradiative Cells

William A. Callahan[1,2], Dudong Feng[3], Zhuomin M. Zhang[3], Eric S. Toberer[2], Andrew J. Ferguson[1], and Eric J. Tervo[1,*]

[1]*National Renewable Energy Laboratory, Golden, CO 80401*
[2]*Department of Physics, Colorado School of Mines, Golden, CO 80401*
[3]*George W. Woodruff School of Mechanical Engineering, Georgia Institute of Technology, Atlanta, GA 30332*
[*]eric.tervo@nrel.gov

Accurate modeling of charge transport and both thermal and luminescent radiation is crucial to the understanding and design of radiative thermal energy converters. Charge carrier dynamics in semiconductors are well-described by the Poisson-drift-diffusion equations, and thermal radiation in emitter/absorber structures can be computed using multilayer fluctuational electrodynamics. These two types of energy flows interact through radiation absorption/luminescence and charge carrier generation/recombination. However, past research has typically only assumed limited interaction, with thermal radiation absorption as an input for charge carrier models to predict device performance. To examine this assumption, we develop a fully-coupled iterative model of charge and radiation transport in semiconductor devices, and we use our model to analyze near-field and far-field GaSb thermophotovoltaic and thermoradiative systems. By comparing our results to past methods that do not consider cross-influences between charge and radiation transport, we find that a fully-coupled approach is necessary to accurately model photon recycling and near-field enhancement of external luminescence. Because these effects can substantially alter device performance, our modeling approach can aid in the design of efficient thermophotovoltaic and thermoradiative systems.

## INTRODUCTION

Solid-state heat engines are attractive alternatives to traditional thermal-fluid power cycles for their ability to serve as compact, scalable electricity generators with no moving parts. Two solid-state systems that utilize thermal radiation, thermophotovoltaic (TPV) and thermoradiative (TR) cells, have the potential to reach high conversion efficiencies [1,2] and could be used with a wide range of source temperatures [3–6]. This makes TPV and TR cells good candidates for a variety of applications, such as solar thermal energy conversion, waste heat recovery, and electricity generation from radioisotopes on spacecraft [7–9]. However, due to the additional considerations of both the thermal emitter and absorber, modeling of TPV and TR systems can be more nuanced than traditional PV systems. Accurate modeling of both charge and radiation transport processes is therefore crucial to the design of high-performance radiative converters.

TPVs generate power through mechanisms similar to traditional photovoltaics, with the distinguishing factor being the utilization of infrared photons from a local thermal emitter instead of solar radiation as illustrated in Fig. 1(a). Photons with energy above the bandgap generate electron-hole pairs, which are separated by the electric field in the depletion region of a p-n diode as shown in Fig. 1(c). Holes (electrons) are collected at contacts on the p-type (n-type) regions, leading to a positive voltage and negative photocurrent indicated in Fig. 1(e) when connected to an external load. TPVs have been widely researched [10–13] with experimental conversion efficiencies reaching about 30% in multiple studies [14–16].

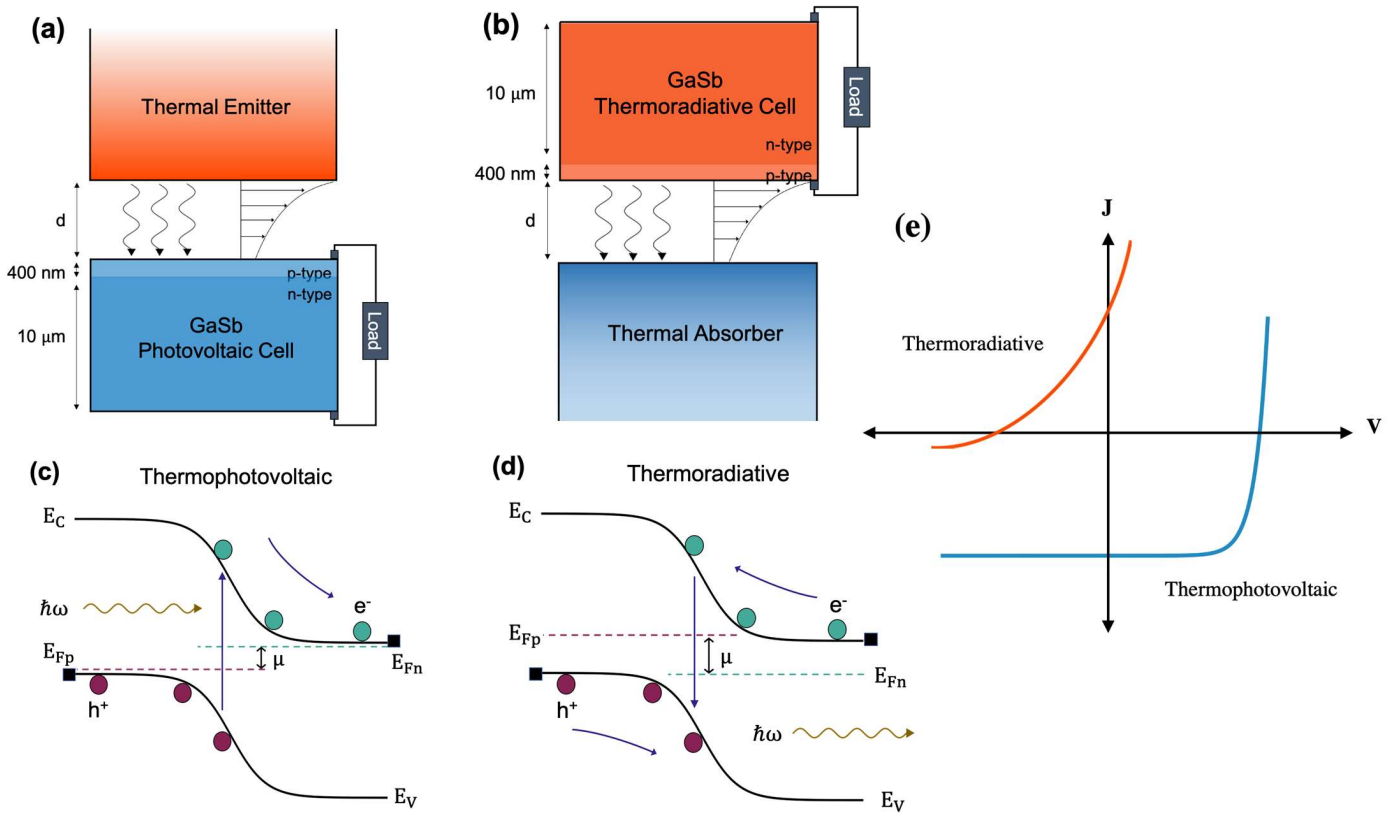

Figure 1: Schematic of GaSb (a) thermophotovoltaic and (b) thermoradiative systems, with their respective band diagrams in (c) and (d). The resulting current/voltage relationships for each system are shown in (e), with thermoradiative cells operating in the second quadrant and thermophotovoltaics operating in the fourth quadrant.

TR cells are a relatively new concept [1,17] and are less intuitive than TPVs. A semiconductor p-n diode in thermal equilibrium with its surroundings is absorbing and emitting photons in equal amounts. If the surroundings are at a lower temperature than the diode, more radiation will be emitted by the semiconductor than absorbed, as illustrated in Fig. 1(b). Correspondingly, the number of minority carriers will be reduced, which results in a split in quasi-Fermi levels opposite that of a TPV as shown in Fig. 1(d). TR devices therefore operate with a negative voltage and a positive photocurrent, which can be thought of as a 'negative illumination.' A representative current-voltage diagram for a TR cell is given in Fig. 1(e). TR cells were only proposed in 2014 [17], but they have been investigated by a number of researchers [18–22] and also could reach high conversion efficiencies.

The performance of TPV and TR systems can be enhanced by utilizing optical modes that contribute to thermal radiation at micro- and nanoscale separation distances between the hot and cold bodies. For gaps smaller than the characteristic thermal wavelength ('near-field' regime), additional evanescent modes due to total internal reflection and surface polaritons enable photon tunneling, which can drastically increase the photon flux [23–35]. This effect can be leveraged to significantly increase the power density of TPV [36–43] and TR [19,44–47] devices compared to 'far-field' operation as a result of the higher radiative generation or recombination rate. Correspondingly, accounting for these near-field effects introduces new challenges in both modeling and experimental design.

As can be inferred from the band structures in Fig. 1, TPV and TR systems follow a very similar device architecture, operate on the same fundamental principles, and can be described with the same equations. In order to accurately understand and design these devices, detailed models compatible with both near-field and far-field operation are needed. The most detailed previous works [48–52] have utilized a fluctuational electrodynamics (FE) framework [23,53] with a scattering matrix method for multilayered media [54] to calculate the spatially resolved net radiative heat flux within the system. From this, the electron-hole pair generation rate is determined at each location within the semiconductor. Finally, a numerical charge transport model is used with the generation rate as an input to calculate how far the semiconductor is moved out of equilibrium and, subsequently, the overall device performance.

This method, which we call the "single-pass" model, has been shown [55] to be insufficient, as it does not capture charge and radiation interactions such as photon recycling and gap-dependent external luminescence. From a fundamental viewpoint, we should expect these interactions to have an effect; the chemical potential of a semiconductor in non-equilibrium alters the net radiation transfer [56,57], which will then change the carrier generation rate. DeSutter *et al.* [55] demonstrated that these effects become particularly significant in the near-field. They showed that external luminescence increases significantly as the separation distance between the two bodies decreases, due to the reciprocity of tunneling modes between emitter and receiver. However, their study focused only on the radiative exchange and did not provide a way to include these effects in conjunction with charge transport. Additionally, for an optically thick semiconductor, non-uniform photon recycling events (photons from radiative recombination that are reabsorbed in the semiconductor and produce electron-hole pairs) play an increasingly important role as the overall photon flux increases. Hence, by neglecting to properly describe the photon transport as both a bidirectional flow between bodies that depends on the charge transport processes and as a spatially non-uniform phenomenon within the cell, the single-pass model may produce errors in the results.

In this work, we propose an iterative model for near-field and far-field TPV and TR systems that includes all spatially varying charge and radiation interactions. Our model couples the Poisson-drift-diffusion (PDD) charge transport equations with a multilayer FE formalism via the quasi-Fermi levels and photon chemical potential, which allows the calculation of self-consistent performance characteristics such as radiative recombination rates, photon recycling, and external luminescence. We compare our iterative model to the single-pass approach detailed by Blandre *et al.* [48], and we find that a fully-coupled iterative model is necessary to accurately predict device performance, especially for nanoscale gaps between the cell and the emitter/absorber.

**MODELING APPROACH**

To calculate the radiative transport in the system, we utilize the FE framework pioneered by Rytov [53]. FE describes thermal emission due to random motion of charges within a material via the fluctuation dissipation theorem [25,35]. These charges, in turn, produce electromagnetic waves that are stochastic in nature. Importantly, this framework can be used to describe radiation in both the near-field and far-field regimes. For the geometries introduced in this paper, we use a scattering matrix formalism for FE as described by Francoeur *et al.* [54]. In this approach, the semiconductor is artificially subdivided into many thin one-dimensional layers, and the thermal emitter or absorber may also be divided into multiple layers as needed. The net spectral radiative heat flux from any layer $s$ to any other layer $l$ is then described as

$$q_{sl}(\omega) = [\Theta(\omega, T_s, \mu_s) - \Theta(\omega, T_l, \mu_l)]\mathcal{T}_{sl}(\omega) \qquad (1)$$

where $\omega$ is the angular frequency, $\Theta(\omega, T, \mu)$ is the mean energy of a Planck oscillator with temperature $T$ and chemical potential $\mu$, and $\mathcal{T}_{sl}(\omega)$ is the spectral transmission coefficient from layer $s$ to layer $l$. The mean energy of a Planck oscillator is defined as

$$\Theta(\omega, T, \mu) = \begin{cases} \dfrac{\hbar\omega}{\exp\left(\dfrac{\hbar\omega - \mu}{k_B T}\right) - 1}, & \hbar\omega \geq E_g \\ \dfrac{\hbar\omega}{\exp\left(\dfrac{\hbar\omega}{k_B T}\right) - 1}, & \hbar\omega < E_g \end{cases} \qquad (2)$$

where $\hbar$ is the reduced Planck constant, $k_B$ is Boltzmann's constant, and $E_g$ is the material's bandgap energy. Note that the inclusion of chemical potential in Eq. (2) allows FE to describe radiative transfer due to both thermal and luminescent photons. For a semiconductor in non-equilibrium, this chemical potential is related to the (spatially varying) quasi-Fermi levels describing the populations of electrons and holes in the device [56,57]:

$$\mu = E_{Fn} - E_{Fp} \qquad (3)$$

where $E_{Fn}$ ($E_{Fp}$) is the electron (hole) quasi-Fermi level. The transmission coefficient in Eq. (1), $\mathcal{T}_{sl}(\omega)$, is an indicator of how radiation travels between layers. It is determined from the difference in probability of transmission from layer $s$ to the front edge of layer $l$, whose position is denoted $z_{l-}$, and from layer $s$ to the back edge of layer $l$, denoted $z_{l+}$. The relation is given by

$$\mathcal{T}_{sl}(\omega) = \int_0^\infty \left[\xi_s(\omega, k_\rho, z_{l-}) - \xi_s(\omega, k_\rho, z_{l+})\right] k_\rho \, dk_\rho \tag{4}$$

where $k_\rho$ is the wavevector parallel to interfaces between layers and $\xi_s(\omega, k_\rho, z_l)$ is the probability of transmission from layer $s$ to a location $z_l$ in layer $l$ given by

$$\xi_s(\omega, k_\rho, z_l) = \frac{k_0^2}{\pi^2} \operatorname{Re} \left\{ i\varepsilon_s''(\omega) \int_{z_{s-}}^{z_{s+}} dz' \begin{bmatrix} g^E_{sl\rho\rho}(k_\rho, z_l, z', \omega) g^{H*}_{sl\theta\rho}(k_\rho, z_l, z', \omega) \\ + g^E_{sl\rho z}(k_\rho, z_l, z', \omega) g^{H*}_{sl\theta z}(k_\rho, z_l, z', \omega) \\ - g^E_{sl\theta\theta}(k_\rho, z_l, z', \omega) g^{H*}_{sl\rho\theta}(k_\rho, z_l, z', \omega) \end{bmatrix} \right\} \tag{5}$$

Here $k_0$ is the wavevector in vacuum, $\varepsilon_s''(\omega)$ is the imaginary part of the relative permittivity $\varepsilon_s(\omega) = \varepsilon_s'(\omega) + i\varepsilon_s''(\omega)$ of layer $s$, and $g^{E(H)}_{sl\alpha\beta}(k_\rho, z_l, z', \omega)$ is the $\alpha\beta$ component of the Weyl representation of the electric (magnetic) dyadic Green's function for a source at $z'$ to a point at $z_l$ in the $(\rho, \theta, z)$ polar coordinate system. These dyadic Green's functions are calculated as described by Francoeur *et al.* [54]. Integration over the parallel component of the wavevector in Eq. (4) is used because this describes both propagating modes for all angles ($k_\rho \leq k_0\sqrt{\varepsilon_s'(\omega)}$) as well as evanescent modes ($k_\rho > k_0\sqrt{\varepsilon_s'(\omega)}$).

By summing the net spectral radiative heat flux, $q_{sl}$, over all layers $s$ in the system and integrating over frequencies of light greater than the bandgap $\omega_g$, we can calculate a net radiative electron-hole pair generation rate for each layer $l$, keeping in mind that this quantity may be positive or negative depending on the temperatures and chemical potentials of the system:

$$G_l = \frac{1}{t_l} \sum_s \int_{\omega_g}^\infty \frac{1}{\hbar\omega} q_{sl}(\omega) \, d\omega \tag{6}$$

where $t_l$ is the thickness of layer $l$. The previous FE equations enable the calculation of net radiative generation rate knowing the materials, geometry, temperatures, and photon chemical potentials of the system.

The photon chemical potentials must be determined from the charge transport dynamics within the cell. We use an open-source Python-based solver of the Poisson-drift-diffusion (PDD) equations called Sesame [58] to capture these effects. Sesame is compatible with one- and two-dimensional systems and can be used to model various optoelectronic devices. Furthermore, Sesame accounts for the entire system; charge transport within the entire device (including the depletion region) is modeled indiscriminately. Most importantly, Sesame allows for custom generation profiles, including those with high spatial resolution, to be assigned. We therefore use the generation term calculated in Eq. (6) and rely on Sesame to solve the following Poisson, continuity, and drift-diffusion equations simultaneously [58]:

$$\nabla \cdot (\varepsilon_{low} \nabla \phi) = -\frac{\sigma}{\varepsilon_0} \quad (7)$$

$$\nabla \cdot \mathbf{J_n} = -e(G - R) \quad (8)$$

$$\nabla \cdot \mathbf{J_p} = e(G - R) \quad (9)$$

$$\mathbf{J_n} = -ev_n n \nabla \phi + eD_n \nabla n \quad (10)$$

$$\mathbf{J_p} = -ev_p p \nabla \phi - eD_p \nabla p \quad (11)$$

Here $\varepsilon_{low}$ is the static (low-frequency) dielectric constant of the material and $\varepsilon_0$ is the permittivity of vacuum; $\phi$ is the electrostatic potential; $\sigma$ is the charge density; $\mathbf{J_n}$ and $\mathbf{J_p}$ are the electron and hole current densities, respectively; $R$ is the recombination rate (including Auger, Shockley-Read-Hall, and surface recombination), which is positive when $\mu > 0$ and negative when $\mu < 0$; and $e$ is the elementary charge. Finally, $n$ and $p$ are the respective concentrations of electrons and holes, with corresponding mobilities $v_n$ and $v_p$. The current densities can also be written as an explicit function of the quasi-Fermi levels [58]:

$$\mathbf{J_n} = ev_n n \nabla E_{Fn} \quad (12)$$

$$\mathbf{J_p} = ev_p p \nabla E_{Fp} \quad (13)$$

We can quickly observe from the PDD and FE equations that they are connected by the photon chemical potentials and the net radiative generation rates, which naturally suggests an iterative approach to solve them. This stands in contrast to the single-pass approach described in the introduction, where the FE equations are used with zero chemical potential to provide a net radiative generation rate for the PDD equations. The computational scheme for both models is broken into the following steps, outlined in Fig. 2:

(1) The system is defined by user-selected materials, geometry, discretized mesh, separation distance, temperatures, and optical properties. The dark (non-illuminated) performance characteristics of the cell are determined with Sesame. An initial chemical potential of $\mu = 0$ is assigned to every point in the system, corresponding to zero luminescence from the cell.

(2) An initial net radiative generation rate $G_l$ for all layers $l$ in the cell is calculated from the FE model (Eq. (6)), utilizing the zero photon chemical potential as an input.

(3) $G_l$ is used to solve the PDD model, and $\mu_l$ is calculated as the difference between quasi-Fermi levels for each layer in the semiconductor (Eq. (3)). At this point, the single-pass model has completed its routine.

(4) The iterative model uses $\mu_l$ to re-calculate $G_l$, and the cycle continues until a pre-defined set of convergence criteria is met for both values. At this point, the iterative model has completed its routine.

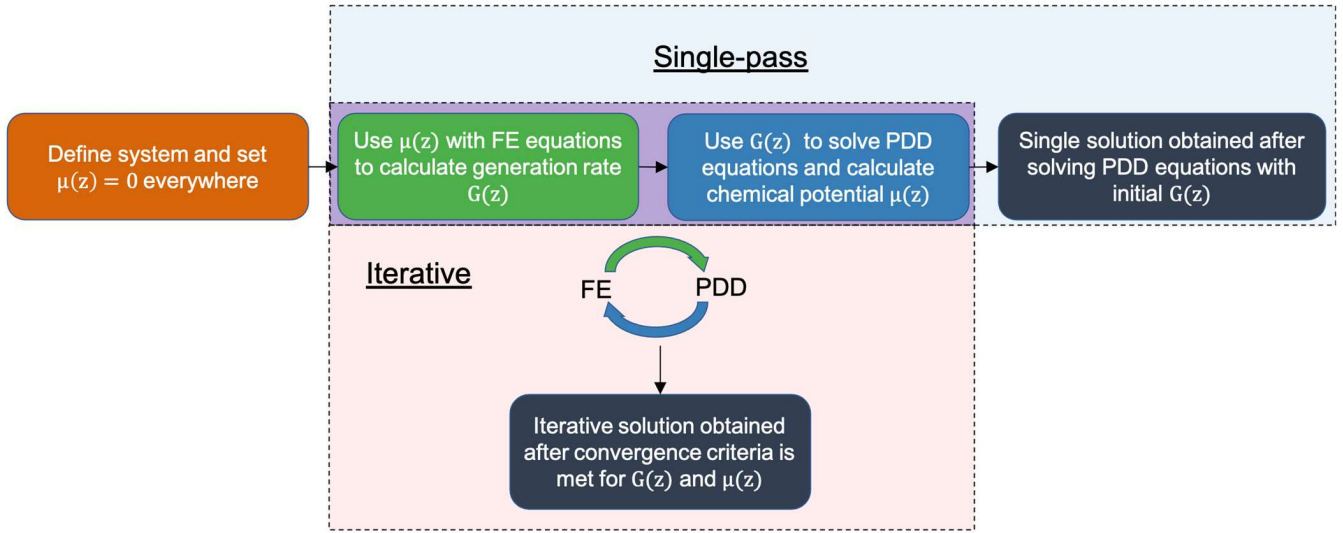

Figure 2: Solution scheme for both single-pass and iterative models

The initialization of a system with a chemical potential $\mu_l = 0$ indicates that the semiconductor is at all points in a state of equilibrium. This means that there can be no nonequilibrium radiative recombination events leading to photon transmission occurring within the first step. In the single-pass method more broadly, because there is no bi-directional communication between the FE and PDD components of the model, this implies that there are no luminescent photons transmitting either back to the emitter or to another location within the cell. Thus, external luminescence and photon recycling events are not captured accurately by the single-pass model. The single-pass model can, however, account for radiative recombination losses of minority carriers using the typical radiative recombination coefficient $B$ and the equation [58]

$$R^{(\text{rad})} = B(np - n_i^2) \qquad (14)$$

where $R^{(\text{rad})}$ is the radiative recombination rate and $n_i$ is the intrinsic carrier concentration. In the single-pass model, this $R^{(\text{rad})}$ is one component of $R$ in addition to any nonradiative losses.

The iterative model relies on convergence criteria for both the chemical potential and the generation rate, so the system necessarily deviates away from the state of zero luminescence. Layers in the system are allowed to have a non-zero chemical potential, and information from a previous step is used to modify future iterations. This functionality, coupled with the use of transmission coefficients between all layers of the system, means that luminescent photons within the cell can contribute to either photon recycling or external luminescence. Accordingly, we can expect this key difference between the models to be reflected in semiconductor characteristics such as current-voltage behavior and carrier concentrations. It is important to note that the major computational expense of both the single-pass and iterative models is the calculation of the transmission coefficients defined by Eq. (4), which can require time on the order of days on a modern personal computer for fine discretization. Convergence of our iterative method, on the other hand, typically requires computational time on the order of minutes.

**THERMOPHOTOVOLTAIC PERFORMANCE**

Calculations were performed for the TPV system illustrated in Fig. 1(a) with inputs selected to match those of Blandre *et al.* [48]. Doping profiles that promote low injection conditions ($N_a = 1\times10^{19}$ cm$^{-3}$, $N_d = 1\times10^{17}$ cm$^{-3}$), moderate minority surface recombination velocity (500 cm s$^{-1}$), and Ohmic contacts were used with a GaSb cell at 300 K. The thermal emitter (2000 K) was described by a Drude model with a plasma frequency $\omega_p = 1.83\times10^{15}$ rad s$^{-1}$ and scattering rate $\Gamma = 2.10\times10^{13}$ rad s$^{-1}$. The system was modeled with a frequency range of $7.53\times10^{14}$ to $37.63\times10^{14}$ rad s$^{-1}$, corresponding to an energy range of 0.496 to 2.5 eV. Dielectric properties for the GaSb cell were calculated from a combination of sources. For radiation with energy below the bandgap of GaSb (0.723 eV), a Drude-Lorentz model was utilized with parameters from [59]. A model formulated by Adachi [60] was used for radiation with energy above the bandgap. Both of these references utilized a doped semiconductor similar to what we present in our model. A length of 1 m was used as the far-field gap, and lengths of 100 nm and 10 nm were used for the near-field gaps. The remaining material parameters are outlined in Table 1. Of particular note, the radiative recombination coefficient for the single-pass model was calculated in the manner outlined by DeSutter *et al.* [55] to match the selected optical properties of GaSb.

Table 1: Simulation parameters for the TPV system.

| Parameter | Value | Reference |
| --- | --- | --- |
| Radiative recombination coefficient [cm$^3$ s$^{-1}$] (single pass only) | $9.6 \times 10^{-11}$ | [55] |
| Auger recombination coefficient [cm$^6$ s$^{-1}$] (electron and hole) | $3 \times 10^{-30}$ | [61] |
| Conduction band density of states [cm$^{-3}$] | $2.81 \times 10^{17}$ | [62] |
| Valence band density of states [cm$^{-3}$] | $6.35 \times 10^{18}$ | [62] |
| Band gap energy [eV] | 0.726 | [48] |
| Electron affinity [eV] | 4.06 | [63] |
| Static relative permittivity | 15.7 | [64] |
| Electron mobility [cm$^2$ V$^{-1}$ s$^{-1}$] | 5544.75 | [65] |
| Hole mobility [cm$^2$ V$^{-1}$ s$^{-1}$] | 322.47 | [65] |
| Electron bulk lifetime [s] | $1.10 \times 10^{-8}$ | [66] |
| Hole bulk lifetime [s] | $3.11 \times 10^{-8}$ | [66] |

For TPV cells, an important indicator of performance is the nonequilibrium band diagram, which also serves as a useful comparison to the model by Blandre *et al.* Fig. 3(a) shows the band diagrams for each separation distance. These band diagrams are shown for the device operating at the voltage of maximum power. This voltage increases as we move from far-field to near-field, corresponding to the increase in injection level within the TPV cell. Fig. 3(a) illustrates two important points: First, the band diagrams demonstrate that our model and that by Blandre *et al.* show good agreement in capturing nonequilibrium conditions of the semiconductor. In fact, our single-pass calculations are able to reproduce their band diagrams nearly identically, providing a validation of our calculations. However, small changes in the quasi-Fermi levels can lead to relatively large changes in the radiation exchange (see Eq. (2)), highlighting the need to look beyond band diagrams to see if the different models lead to different results. Second, we see behavior in the photon chemical potential that is contrary to a common assumption when modeling TPV systems, as shown in Fig. 3(b). Typically, the chemical potential (the difference between the quasi-Fermi levels) is assumed to be equal to the applied voltage (that is, $\mu = qV$) and uniform throughout the cell [55,67–69]. We see that both of these assertions are not necessarily true. Our band diagrams show that the applied voltage assumption underestimates the actual chemical potential in the TPV cell. This discrepancy is due to high-injection effects which increase the minority carrier concentrations and cause the difference in quasi-Fermi levels to exceed the applied voltage. Additionally, the chemical potential is clearly spatially non-uniform, especially at the junction and deep in the bulk of the cell.

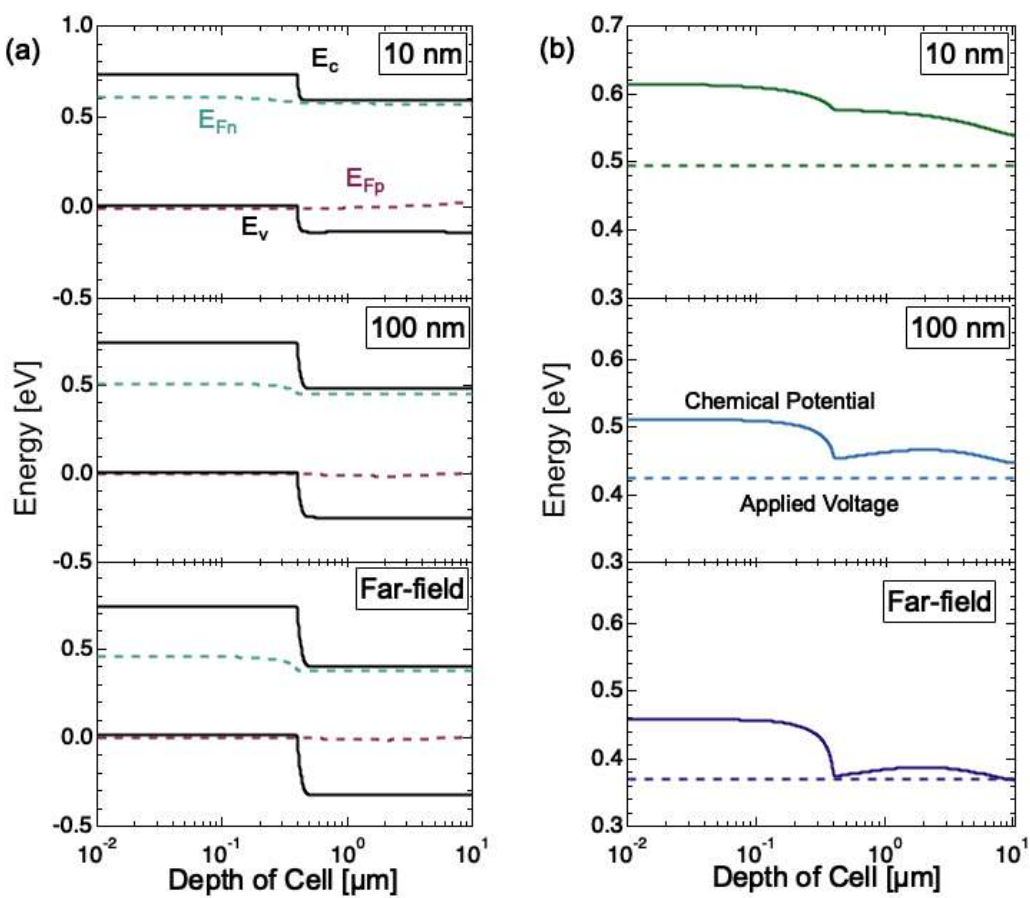

Figure 3: (a) Band diagrams and (b) spatial photon chemical potentials for a TPV system at 10 nm, 100 nm, and far-field separation distances. Biases of 0.494V, 0.425V, and 0.37V were respectively applied, shown by the dashed line in (b), corresponding to the maximum power point for each separation distance.

Although the band diagrams provide useful information about the quasi-Fermi levels, they do not directly describe figures of merit related to device performance. For this we turn to the current-voltage relationship, which we show in Fig. 4(a) with electricity generation moved to the first quadrant. Qualitatively, we see that for far-field and 100 nm separation distances, the iterative model predicts current densities that are slightly higher than the single-pass model. However, for 10 nm the opposite is true, and the single-pass model predicts significantly higher current densities than the iterative model. These differences in current can translate to substantial differences in output power, shown in Fig 4(b). At 100 nm and far-field separation distances, the output power calculated by the single pass model is about 3% below that calculated by the iterative model (6.17 W cm$^{-2}$ vs. 6.37 W cm$^{-2}$ and 0.51 W cm$^{-2}$ vs. 0.53 W cm$^{-2}$, respectively). For the 10 nm separation gap, the single-pass model predicts an output power more than 10% higher than the iterative model (97.6 W cm$^{-2}$ vs 88.5 W cm$^{-2}$).

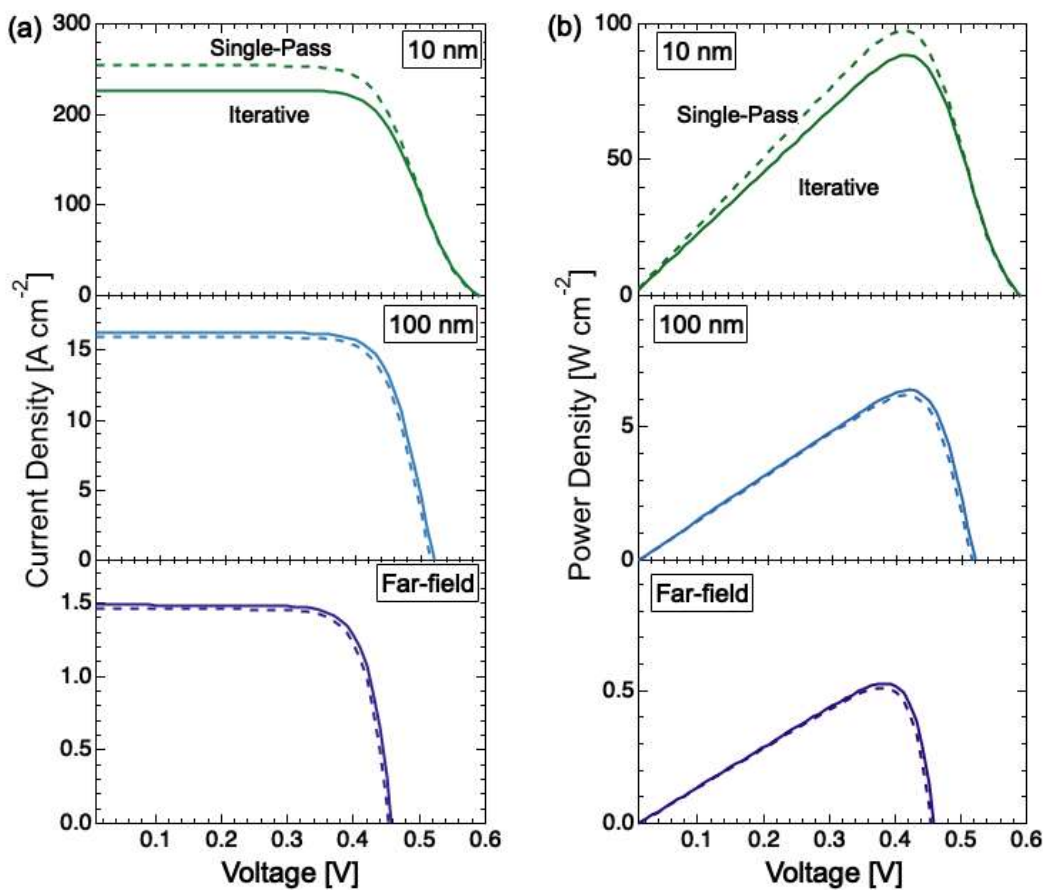

Figure 4: Plots of (a) current density vs. voltage and (b) power density vs. voltage for a TPV system with three different separation distances for both the single-pass and iterative models.

Efficiency in the context of thermophotovoltaics is defined as the power output of the device divided by the full-spectrum radiative heat flux (i.e., energies above and below bandgap). Due to the unoptimized design/materials used in this demonstration, these devices operate at very low efficiencies. For the iterative model, the efficiency values are 4.79×10$^{-3}$%, 4.30×10$^{-3}$%, and 2.72×10$^{-3}$% for 10 nm, 100 nm, and far-field, respectively. For the single-pass model, these values are 4.62×10$^{-3}$%, 4.15×10$^{-3}$%, and 2.63×10$^{-3}$%. Generally, the iterative model predicts slightly higher efficiencies than the single-pass model. The efficiency could be drastically improved with appropriate materials selection and cell design [70]. Features like a back surface reflector can increase photon utilization, and cladding layers can reduce surface recombination events. In the following paragraphs, we explore the reasons why the single-pass and iterative models are different.

The explanation of performance as a function of separation distance can be traced back to the behaviors of photons and of charge carriers. We begin by focusing on photon transport. Internal luminescence in the iterative model is a blanket term for all luminescence within the cell and can result in two different events: photon recycling, in which radiative recombination events produce a photon that fails to leave the TPV cell and instead generates an additional electron-hole pair; or external luminescence, in which the resulting photon leaves the TPV cell without producing another electron-hole pair. Photon recycling is considered beneficial, as it effectively prolongs carrier lifetimes, thereby increasing the probability that charge carriers will be collected and contribute to photocurrent. External luminescence is considered a loss mechanism, for the opposite reason. In the single-pass model, there is no distinction between internal and external luminescence; all radiative recombination events calculated with the coefficient as shown in Eq. (14) are considered to be loss mechanisms. In the iterative model, the local internal luminescence is calculated with

$$\gamma_{i,l} = \frac{1}{t_l} \int_{\omega_g}^{\infty} \sum_{s} \frac{d\omega}{\hbar\omega} \Theta(\omega, T_l, \mu_l) \mathcal{T}_{sl} \qquad (15)$$

The calculation for local external luminescence is very similar, but instead of summing over all layers in the system, we only consider transmission to those that are not in the cell (i.e., regions where photons have escaped):

$$\gamma_{e,l} = \frac{1}{t_l} \int_{\omega_g}^{\infty} \sum_{s \notin cell} \frac{d\omega}{\hbar\omega} \Theta(\omega, T_l, \mu_l) \mathcal{T}_{sl} \qquad (16)$$

We can calculate the photon recycling as the number of emitted above bandgap photons that remain in the cell. In this case, we portray the photon recycling as a "destination" effect – that is, the number of photons that are absorbed at a particular location in the cell that come from all other parts of the cell. We can write this as

$$\gamma_{r,l} = \frac{1}{t_l} \int_{\omega_g}^{\infty} \sum_{s \neq l} \frac{d\omega}{\hbar\omega} \Theta(\omega, T_s, \mu_s) \mathcal{T}_{sl} \qquad (17)$$

We can see the comparison of the two key luminescence metrics in Fig. 5. The external luminescence is shown in Fig. 5(a), comparing the iterative model (Eq. (16)) to the radiative recombination rate predicted by the single-pass model (Eq. (14)). In both the far-field and 100 nm cases, the single-pass model yields a radiative recombination rate that is higher than the external luminescence of the iterative model. As the gap is further reduced to 10 nm, this behavior changes. At the very front of the cell, the external luminescence in the iterative model is an order of magnitude greater than the radiative recombination in the single-pass model. This means that the iterative model predicts a large number of photons emanating from the front of the cell and contributing to loss of current. A mathematical explanation for this is that the transmission coefficients between layers are symmetric: decreasing separation distance dramatically increases the photon flux from emitter to absorber, but accordingly also increases the flux from the absorber back to the emitter. The photons leaving the cell are created by charge carriers that are recombining and not contributing to current collection, so we consequently see a large decrease in current density in the iterative model compared to the single-pass model. It is important to note that the external luminescence predicted by Eq. (16) for the single-pass model (instead of Eq. (14), which is plotted in Fig. 5(a) for the single-pass model) is negligibly small, because $\mu_l = 0$ everywhere in the single-pass model and the temperature of the cell is too low for substantial thermal emission above the bandgap.

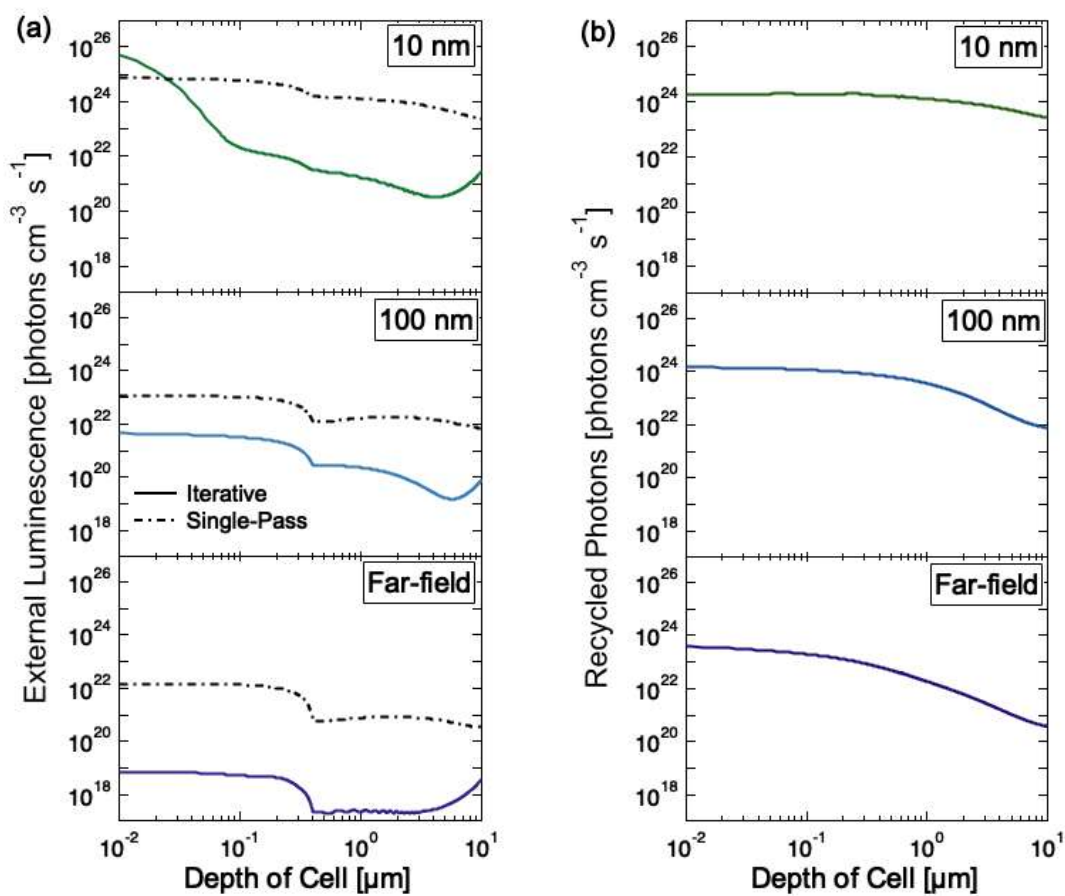

Figure 5: (a) External luminescence (iterative) and radiative recombination rate (single-pass) as a function of position in the TPV cell at the maximum power point voltage. (b) Local absorption of recycled photons for the same case as (a) from the iterative model.

By comparing the external luminescence to the total recombination losses (both radiative and non-radiative), we can further use Fig. 5(a) to explain the differences in performance we see in the JV curves. Total recombination losses can be calculated directly from Sesame. In the iterative model at 10 nm separation distance, radiative recombination accounts for more than 61% of the total recombination losses of minority carriers; at 100 nm and far-field gaps, however, radiative recombination is less than one percent of total recombination losses (0.88% and 0.12%, respectively). For the single-pass model the radiative portion of recombination losses for 10 nm, 100 nm, and far-field gaps are 32.69%, 26.3%, and 29.79%, respectively. This means that although the external luminescence in both the far-field and 100 nm cases is lower for the iterative model than in the single-pass model, the non-radiative losses are substantially greater and dictate overall performance. The result of this are current-voltage curves that are similar between the two models for far-field and 100 nm gaps. In the 10 nm case, a significant portion of the total losses in the iterative model are radiative, which is reflected in the current-voltage curve. This is a direct result of the dramatic increase in photon flux, especially at the front of the cell.

The local absorption of recycled photons in the iterative model is shown in Fig 5(b). Recall that in the single-pass model, the photon recycling is zero because all emitted photons from radiative recombination predicted by Eq. (14) are considered to be lost and the temperature of the cell is too low for substantial thermal emission above the bandgap. We see that as the separation distance decreases, the magnitude of photon recycling increases significantly, which is associated with the larger photon chemical potentials and greater internal luminescence for smaller separation distances. Additionally, the large number of recycled photons for the 100 nm and far-field cases help to reduce the overall impact of radiative recombination in the iterative model, as described in the previous paragraph. Comparing to the external luminescence in Fig 5(a), we note that in the far-field and 100 nm cases the number of photon recycling events is much larger than the number of photons lost to external luminescence, which helps to explain the larger currents predicted by the iterative model for these cases. Furthermore, at 10 nm, the iterative model predicts much higher external luminescent losses than photon recycling events, which is consistent with the lower current predicted by the iterative model.

The effects of external luminescence and photon recycling can also be seen in the calculated carrier concentrations for each separation distance, shown in Fig. 6. Fig 6(a) shows the carrier concentration plotted against position within the cell. Given the magnitude of carrier concentrations and the use of the log scale, we see what appear to be visually small differences. To better distinguish between the outputs of each model, in Fig 6(b) we plot the relative error of the single-pass model compared to the iterative model. From this, we can make several observations. First, the single-pass model dramatically under-predicts the concentration of minority carriers in the bulk of the cell at all separation distances, which is consistent with the significant number of recycled photons absorbed here as predicted by the iterative model. Second, at 10 nm the single-pass method over-predicts the number of electrons at the front of the cell. This is consistent with the results in Fig. 5(a), which shows the iterative model predicting very large external luminescent losses in this region. The transmission coefficients are very large in this part of the cell, and the iterative model captures the ability of luminescent photons from the cell to tunnel back to the emitter just as thermal photons can tunnel from the emitter to the cell. These losses result in fewer minority carriers because of the effectively low radiative lifetimes at the front of the cell.

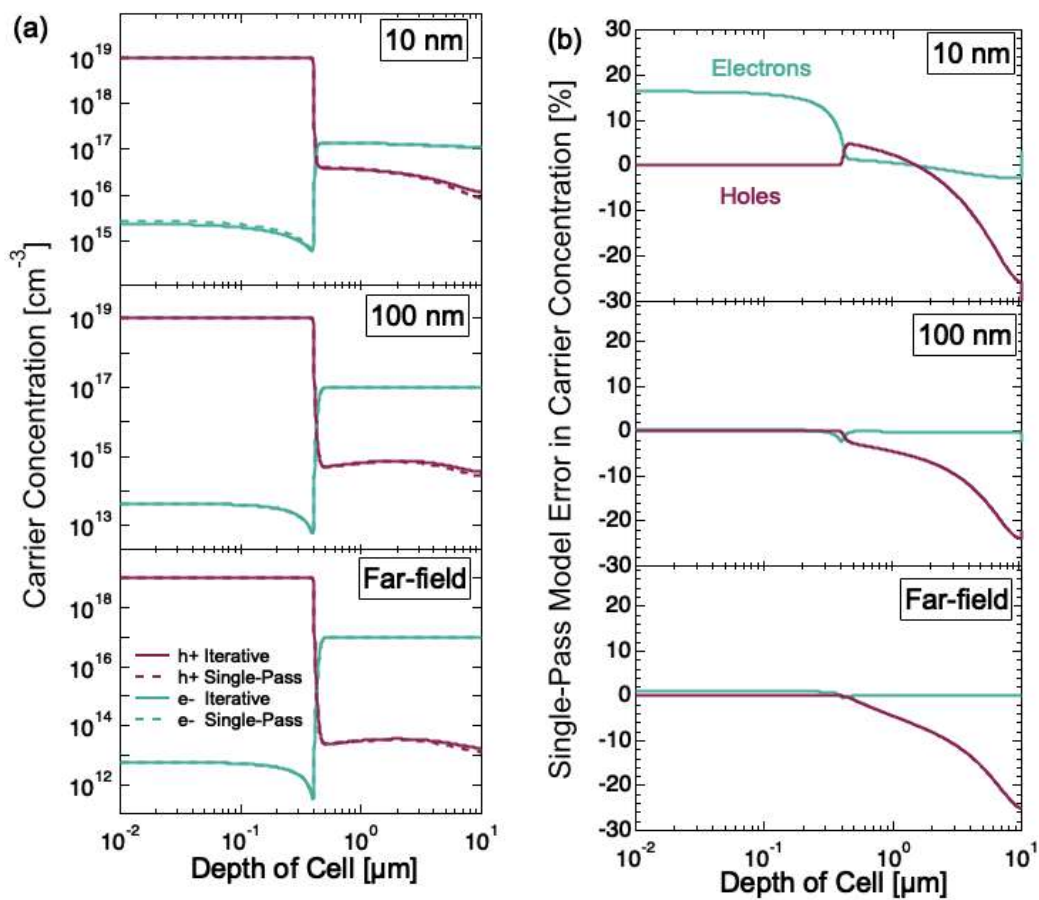

Figure 6: (a) Charge carrier concentration as a function of position in cell. (b) Relative error in carrier concentration predicted by the single-pass model with respect to the iterative model.

**THERMORADIATIVE PERFORMANCE**

The TR system shown in Fig. 1(b) was modeled with the same input parameters as the TPV case except for the system temperatures and the acceptor doping. The GaSb cell temperature is set to 900 K to stay below its melting point, and the Drude absorber is set to 300 K. In practice, elevated temperatures would cause some inputs for GaSb to change (e.g. band gap, carrier mobilities and lifetimes, etc.), but our system does not reflect these temperature dependencies. The acceptor doping has been reduced to $10^{18}$ cm$^{-3}$ to avoid substantial degeneracy. Our intention with these inputs is not to design an optimized TR system but to see whether the proposed iterative model will predict different performance than the single-pass model for a TR device just as it does for the TPV system.

We again use the current-voltage relationship as a primary method of comparison, shown in Fig. 7 with electricity generation moved to the first quadrant for the 10 nm separation distance. The single-pass model once again predicts a better absolute performance than the iterative model for this separation distance, with a higher maximum power output of about 33% (2.09 mW cm$^{-2}$ for single-pass vs. 1.57 mW cm$^{-2}$ for iterative). These low power densities result from the use of GaSb as the TR cell, as its bandgap is much too high to achieve significant power output at achievable temperatures (i.e. $E_g \gg k_B T$). The 100 nm and far-field separation distances are not shown, as these have even lower power outputs due to the reduced photon exchange compared to the 10 nm case. Nevertheless, the substantial differences between the iterative and single-pass models here indicate that the interactions between charge and radiation transport are significant for TR devices just as they are for TPV systems.

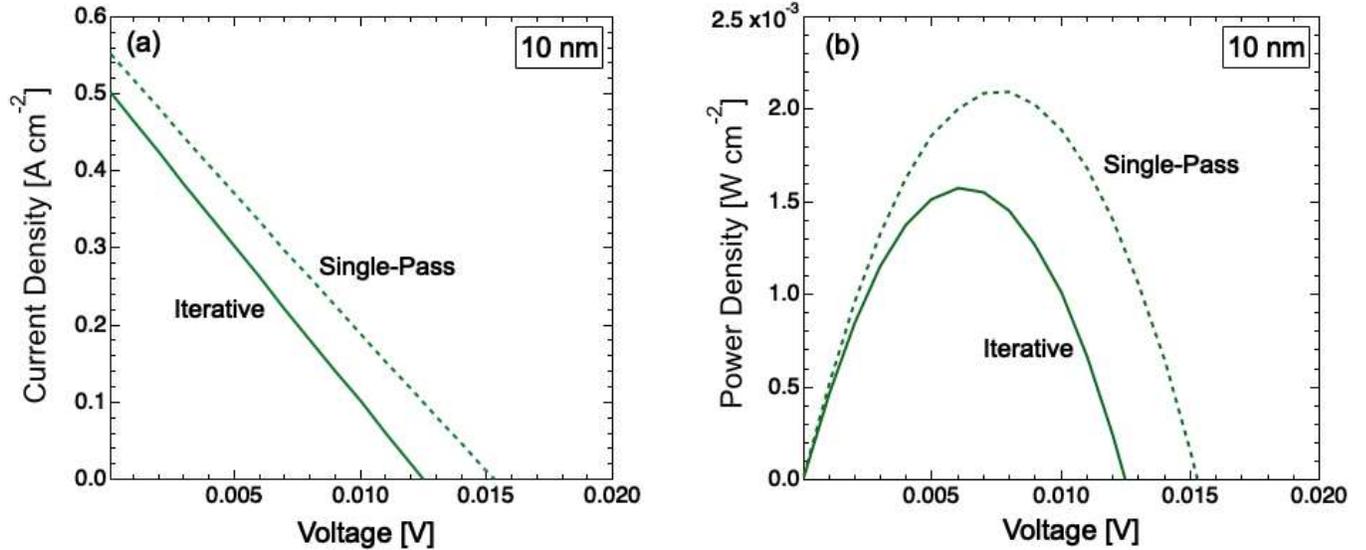

Figure 7: (a) Current density – voltage and (b) power density – voltage relationships for a GaSb thermoradiative cell at 900 K emitting to a 300 K thermal absorber with a separation distance of 10 nm.

To explain the difference between the performance of the two models, we look again at the luminescence in this system. The external luminescence and photon recycling are calculated for the iterative model with Eqs. (16) and (17) as with the TPV system. Since the temperature of the cell is higher in this case, there will be more thermal emission above the bandgap, and we can also use these equations for the single-pass model with $\mu = 0$ everywhere. The radiative recombination coefficient is set to zero for the single-pass model, because radiative generation losses in a TR cell can only result from an absorbed photon and are therefore not necessarily proportional to $(np - n_i^2)$ as given by Eq. (14). The current in a TR cell is driven by radiative recombination. This means that in contrast to a TPV system, external luminescence is a desirable effect. Photons leaving the system result from charge carriers recombining to complete the circuit, and photons that are recycled cause unwanted electron-hole pair generation.

Plotted in Fig. 8(a) is the external luminescence comparison between iterative and single-pass models for this TR system. Similar to the TPV system at the same separation distance, we see an increase in external luminescence at the front of the cell in both single-pass and iterative models. The local absorption of recycled photons for the system is plotted in Fig. 8(b), with the same axis scales as Fig. 8(a) for comparison. Both plots in Fig. 8 show insets with a linear scale to demonstrate that at all points, the single-pass model predicts much more external luminescence and photon recycling than the iterative model. We can explain this by looking directly at how these values are calculated. Because the iterative model considers the chemical potential, and because a TR cell operates with inverted quasi-Fermi levels compared to a TPV cell (i.e., the quasi-Fermi level for holes is at a higher energy than that for electrons), the resulting negative photon chemical potential will result in fewer photons emitted from the cell when compared to single-pass calculations that omit the chemical potential. This demonstrates that considering the photon chemical potential and the interactions between charge and radiation transport processes are critical to avoid overpredicting the performance of TR cells.

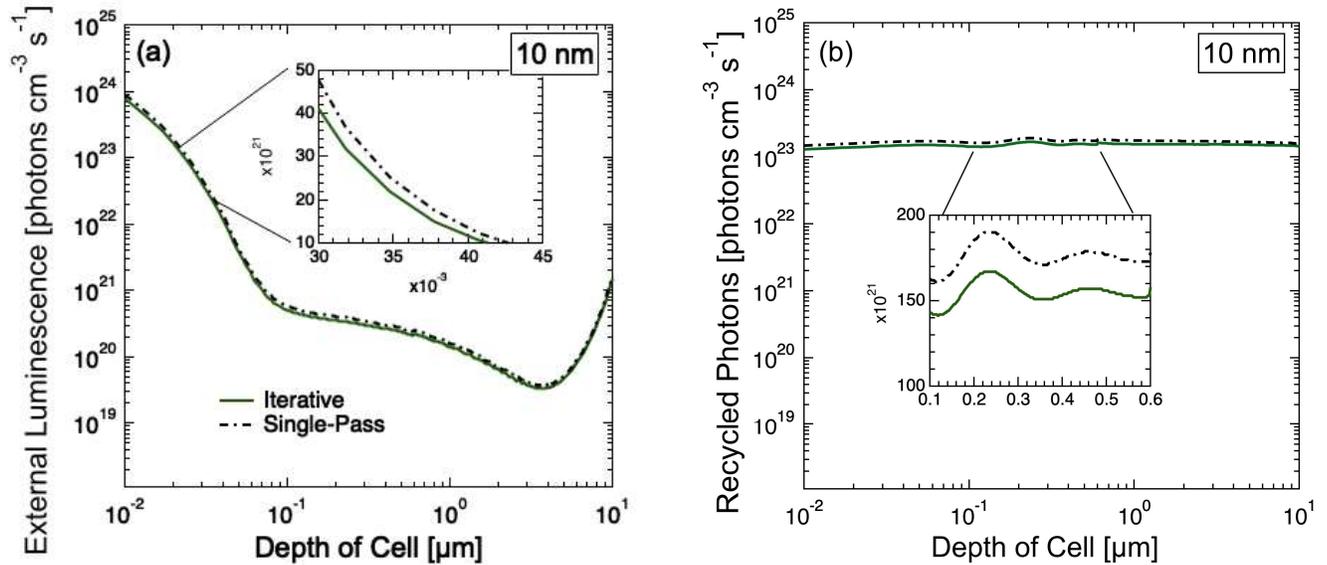

Figure 8: (a) External luminescence as a function of position in the TR cell at the maximum power point voltage. (b) Local absorption of recycled photons for the same case as (a).

Preprint of published article: *Physical Review Applied* **15**, 054035 (2021)
https://doi.org/10.1103/PhysRevApplied.15.054035## CONCLUSIONS

We presented a fully-coupled iterative model of charge and radiative transport in TPV and TR systems, compatible with both near-field and far-field regimes. To this, we compared a model from previous literature that utilizes the same foundational equations but does not consider bidirectional interactions between light and charge transport in the system. We have shown that an iterative and coupled approach to these interactions is necessary for accurate performance prediction in both TPV and TR systems. Even when the two models have reasonable agreement in the current-voltage characteristics, the iterative model predicts significant differences in other cell properties such as net radiative recombination rate and carrier concentration.

In this work we considered a very specific material system (GaSb cell and a Drude emitter/absorber) at two separate temperatures and three separation distances. This system was chosen for an initial comparison of the models due to its simplicity and well-characterized material properties, but there are a great many other possible materials, cell architectures, temperatures, and separation distances that are possible. Future work should investigate comparisons of the two models for other types of systems, such as thin-film TPVs with rear reflectors or low bandgap TR cells. The PDD solver used in this work, Sesame, is compatible with two-dimensional geometries, but here we only considered one-dimensional systems. Our approach could be extended to two-dimensional systems to model converters that include more complex two-dimensional features. Additional work can be done to improve the resource efficiency of the model; specifically, calculation of the transmission coefficients act as a computational bottleneck and should be optimized.

Our iterative model should provide researchers with the ability to more accurately predict the performance and spatial transport characteristics of radiative thermal energy converters. It can be applied to better design TPV and TR systems, especially when they operate in the near-field, but it also could be applied to other types of solid-state converters such as electroluminescent refrigerators. This should enable a better-informed design process that ultimately leads to more efficient and higher power density solid-state conversion systems.

## ACKNOWLEDGEMENTS

This work was authored in part by the National Renewable Energy Laboratory (NREL), operated by Alliance for Sustainable Energy, LLC, for the U.S. Department of Energy (DOE) under Contract No. DE-AC36-08GO28308. This work was supported by the Laboratory Directed Research and Development (LDRD) Program at NREL. D.F. and Z.M.Z. would like to thank support from the U.S. Department of Energy, Office of Science, Basic Energy Sciences (DE-SC0018369). The views expressed in the article do not necessarily represent the views of the DOE or the U.S. Government. The U.S. Government retains and the publisher, by accepting the article for publication, acknowledges that the U.S. Government retains a nonexclusive, paid-up, irrevocable, worldwide license to publish or reproduce the published form of this work, or allow others to do so, for U.S. Government purposes.